\documentstyle[floats,prl,aps]{revtex}
\input epsf
\begin{document}

\newcommand{\vertsp}{\vphantom{\displaystyle{\dot a \over a}}}
\newcommand{\se}{{(0)}}
\newcommand{\ve}{{(1)}}
\newcommand{\te}{{(2)}}
\newcommand{\nnu}{\nu}
\newcommand{\Spy}[3]{\, {}_{#1}^{\vphantom{#3}} Y_{#2}^{#3}}
\newcommand{\Gm}[3]{\, {}_{#1}^{\vphantom{#3}} G_{#2}^{#3}}
\newcommand{\Spin}[4]{\, {}_{#2}^{\vphantom{#4}} {#1}_{#3}^{#4}}
\newcommand{\scpot}{{\cal V}}
\newcommand{\tl}{\tilde}
\newcommand{\bm}{\boldmath}
\newcommand{\MNRAS}{Mon. Not. Roy. Astron. Soc.}
\def\bi#1{\hbox{\boldmath{$#1$}}}

\renewcommand{\ell}{l}
\renewcommand{\topfraction}{1.0}
\renewcommand{\bottomfraction}{1.0}
\renewcommand{\textfraction}{0.00}
\renewcommand{\dbltopfraction}{1.0}

\draft

\title{Lensing of the CMB: Non Gaussian aspects}

\author{Matias Zaldarriaga\cite{matiasemail}}
\address{Institute for Advanced Studies, School of Natural Sciences,
Princeton, NJ 08540}

\maketitle

\begin{abstract}

We study the generation of CMB anisotropies by gravitational lensing
on small angular scales. We show these fluctuations are not Gaussian.
We prove that the power spectrum of the tail of
the CMB anisotropies on small angular scales directly gives the
power spectrum of the deflection angle. We show that the generated
power on small scales is correlated with the large scale gradient. 
The cross
correlation between large scale gradient and small scale power can
be used to test the hypothesis that the extra power is indeed
generated by lensing.  
We compute the three and four point function of the
 temperature in the small angle limit. 
We relate the non-Gaussian aspects presented in
this paper as well as those in our previous studies of the
lensing effects on large scales to the three 
and four point functions. We interpret the statistics proposed 
in terms of different configurations of the four
point function and  show how they relate to the statistic that
maximizes the $S/N$. 

\end{abstract}

\pacs{PACS numbers: 98.80.Es,95.85.Bh,98.35.Ce,98.70.Vc  \hfill}

\section{Introduction}

The anisotropies in the Cosmic Microwave Background (CMB) are
thought to contain detailed information about the underlying
cosmological model. In conventional models the anisotropies on
most angular scales were created at the last scattering surface,
at a redshift of $z\sim 1000$. At these early times the evolution
of perturbations can be calculated accurately with linear theory.
The calculation of theoretical predictions is almost
straightforward, thus detailed observations of the microwave sky
can, at least in principle, greatly constrain the cosmological
model. We expect to be able to measure many of the parameters of
the cosmological model with percent accuracy \cite{parameters}. 

There are several physical processes that imprint anisotropies on
the CMB after decoupling. Some of them will degrade our ability to
learn about cosmology, like foreground emission from our
galaxy. Others will allow us to constrain processes that happen
after decoupling and help us understand the evolution of our
universe. For example the reionization of hydrogen by the
ultraviolet light from the first generation of objects leaves a
distinct mark in the polarization of the CMB \cite{zalrei},
Sunyaev-Zeldovich emission from hot gas along the line of sight
creates temperature anisotropies and the large scale structure
(LSS) of the universe deflects the CMB photons, lensing the
anisotropies \cite{bernardeau,goldsper,psdmletter,longlens,uroslens}.

When studying the lensing effect produced by the large scale
structure of the universe  we are trying to detect lensing
produced by random mass fluctuations, the LSS, on a random
background image, the CMB. The characteristics of the lensing
effect depends on the relative size of the coherence lengths of
these two random fields. In \cite{psdmletter,longlens} we studied
the limit of a rapidly fluctuating CMB background being lensed by
a slowly varying mass distribution. This is the appropriate limit
to recover the power spectrum of the projected mass density  on
scales much larger than the coherence length of the CMB, $\xi\sim
0.15^o$. In this paper we study the opposite limit, the generation
of power on scales much smaller than $\xi$.

The lensing effect is expected to be the dominant non primordial
contribution to the CMB anisotropies on small scales ($l \sim 3000$). 
It has been shown \cite{metcalf} that an accurate 
determination of the power generated by lensing can help break 
some of the parameter degeneracies in the CMB. Interferomentric
observations of the anisotropies such as those that will be
carried out by CBI \cite{cbi} are designed to make measurements at
these angular scales. 

To be able to use the observed power on small scales to break the
degeneracies in the parameters one must be sure that one is observing
the lensing signal. 
We will show that the small scale power generated by gravitational
lensing has a very definite signature, it is correlated with the
large scale gradient. Regions of the sky where the large scale
gradient is larger will have more small scale power. The physical
effect can be understood easily in the case of a cluster of
galaxies lensing a smooth CMB gradient.

We will also compute the general three and four point function of the
temperature field induced by lensing and show that both the
statistic discussed in this paper and those proposed in
\cite{longlens} are particular subsets of the possible
configurations of the four point function. We will construct the
statistic that maximizes the signal to noise ratio.

\section{Generation of power on small scales}

The measured temperature field $T({\bi \theta})$ can be expressed
in terms of the unlensed CMB field at the last scattering surface
$\tl T({\bi \theta})$ and the deflection angle of the CMB photons
$\delta {\bi \theta}$,
\begin{eqnarray}\label{expansion}
T({\bi \theta})&=&\tl T({\bi \theta}+\delta {\bi \theta})
\nonumber \\ &\approx& \tl T ({\bi \theta}) +\delta {\bi
\theta}\cdot \nabla \tl T({\bi \theta}) + {1 \over 2} \delta\theta_i
\delta\theta_j \partial_{ij}\tl T ({\bi \theta}).
\end{eqnarray}
In this paper we are interested in the effect of modes of the
deflection angle of spatial wavelength  much smaller than that of the
unlensed CMB.

\subsection{A toy example}

To understand the physics we will first consider the lensing
induced by a cluster of galaxies. We will discuss a very simplified
example in this paper, the reader interested in the detectability of
the effect for real cluster should look at reference \cite{uroszalcluster}. 
In most cases a  cluster will
subtend a few arcminutes on the sky. Over such a  scale the
primary anisotropies are expected to have negligible power so we
will treat the unlensed temperature field as a pure gradient.
Without loss of generality we will take the gradient to be along
the $y$ axis with an amplitude $\tl T_{yo}$. The observed
temperature becomes,
\begin{equation}\label{tunlensedcl}
  T(\bi \theta)=\tl T_{y0} (\theta_y + \delta \theta_y).
\end{equation}
For a spherically symmetric cluster at the origin the deflection
is of the form,
\begin{equation}\label{defcluster}
  \delta \bi \theta=-\delta\theta {\bi\theta \over \theta}.
\end{equation}
For a singular isothermal sphere $\delta\theta=4\pi (\sigma_v/c)^2
D_{LS}/D_{OS}$, with $\sigma_v$ the cluster velocity dispersion,
$D_{LS}$ the distance from the lens to the source and $D_{OS}$
that between the observer and the source. In an Einstein DeSitter
universe $D_{LS}/D_{OS}\approx 1/\sqrt{1+z_L}$, with $z_L$ the
redshift of the cluster and where we have assumed that $z_L$ is
much smaller than the redshift of recombination.  A singular
isothermal profile will be a good approximation for a cluster up
to some maximum radius after which the deflection will fall as
$1/\theta$. In the central part of the cluster there might be a
core. The typical value for $\delta \theta$ is a fraction of an
arcminute.

\begin{figure*}
\begin{center}
\leavevmode \epsfxsize=6.0in \epsfbox{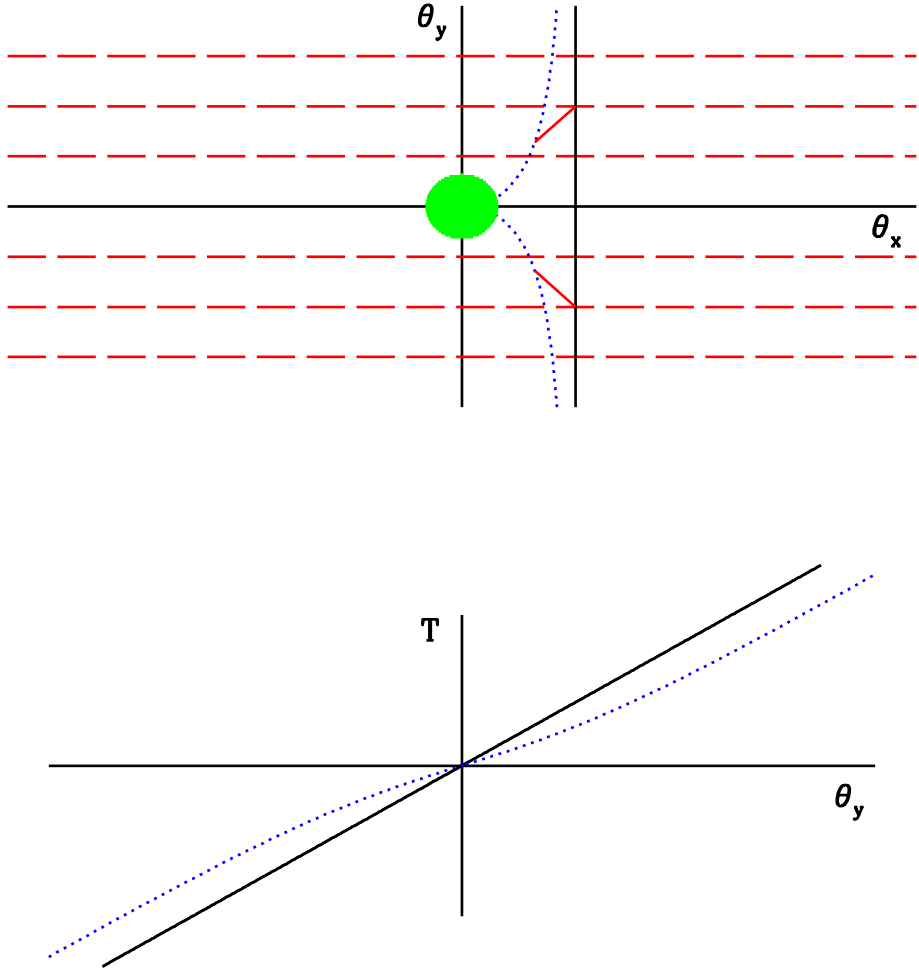}
\end{center}
\caption{In the upper panel we show a cluster lensing a background
gradient. The bottom panel shows the temperature measured for a
fixed $\theta_x$ as a function of $\theta_y$ in the presence and
absence of the cluster. Points with $\theta_y>0$ get deflected to
a smaller $\theta_y$ in the lens plane and thus for a positive
gradient they will have a lower temperature in the lensed example
than in the unlensed one. The opposite is true if $\theta_y<0$.}
\label{sketch}
\end{figure*}

\begin{figure*}
\begin{center}
\leavevmode \epsfxsize=6.0in \epsfbox{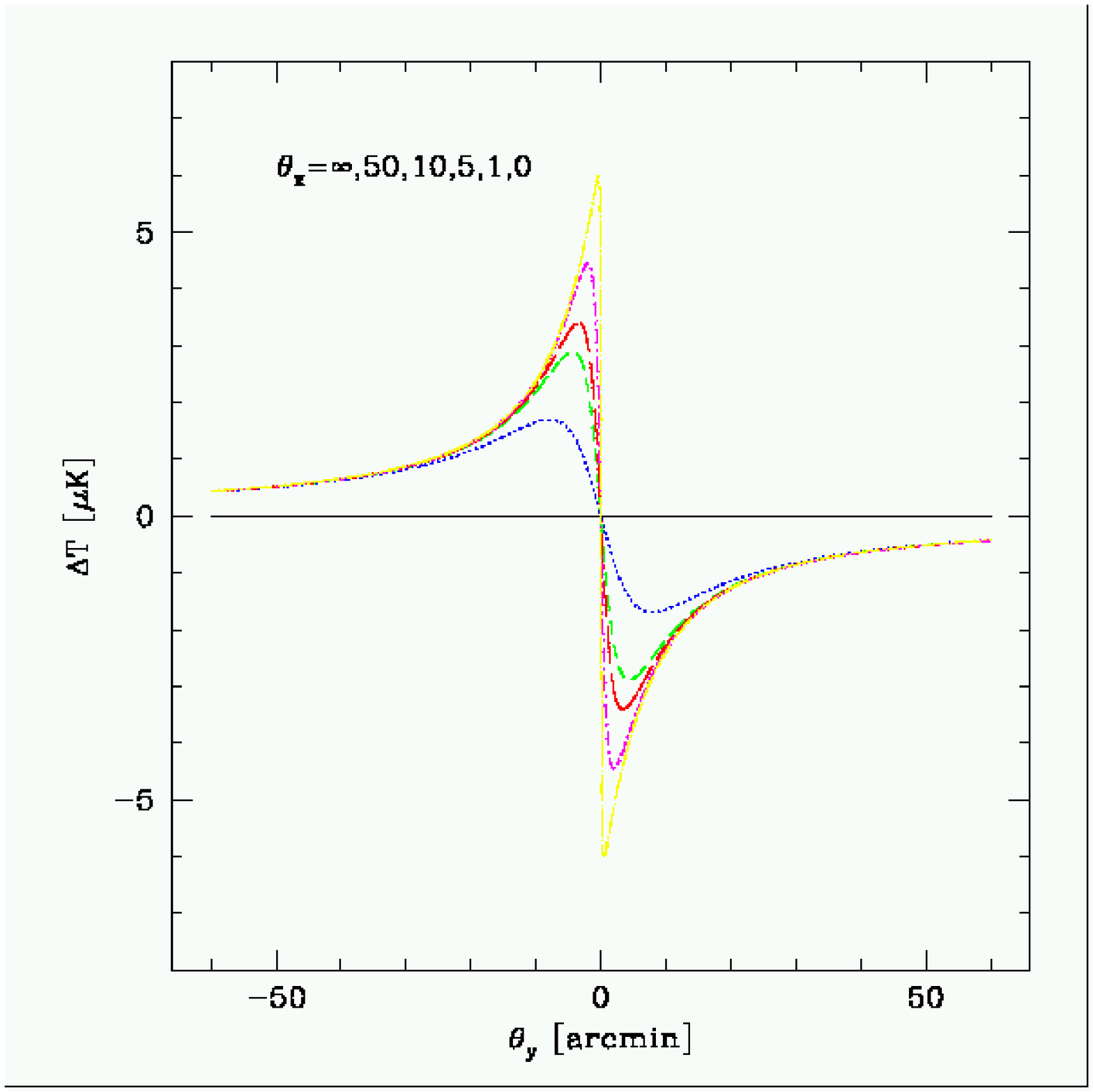}
\end{center}
\caption{Temperature profile of a CMB gradient lensed by a
cluster. We took $T_{y0}\delta\theta= 13 \mu K$. The gradient part
has been subtracted out for clarity. The cluster profile was
cut-off at $4\ {\rm arcmin}$.}\label{clusterwiggle}
\end{figure*}

The effect of lensing can be understood by looking at figure
\ref{sketch}. We focus on the temperature as a function of
$\theta_y$ for a fixed $\theta_x$. In the absence of lensing we
would observe the gradient. On the other hand the cluster will
deflect the light rays so that for $\theta_y > 0$ the rays are
coming from a lower value of $\theta_y$ in the last scattering
surface. If the gradient is positive this implies that for
$\theta_y > 0$ in the presence of the cluster we would observe a
lower temperature than what would be observed if the cluster was
not there. The opposite is true for $\theta_y < 0$. Far away from
the cluster the lensed temperature should coincide again with the
gradient. Thus the cluster creates a wiggle on top of the large
scale gradient. This effect is shown in figure
\ref{clusterwiggle}. The size of this wiggle is
$T_{y0}\delta\theta$, thus it is proportional to the size of the
gradient and the deflection angle. For a cluster we can use the
determined deflection  to infer the mass and some information
about the cluster profile.

Note that the proposed signal is directly sensitive to the
deflection angle and not the shear. This is the only method that
has this property. It is usually argued that we can never measure
the deflection angle in a lensing system because we do not know
the original position of the background image. In this case we
avoid this argument because we do know what the background image
is, it is a gradient which we can measure on larger scales  than
the cluster. The other important point is that the effect is
proportional to the gradient $T_{y0}$, this is the signature we
will use in what follows to identify the lensing effect by large
scale structure in the limit we are studying.

\subsection{Large scale structure}

For simplicity we will work in the small angle approximation so
that we can expand the temperature field in Fourier modes instead
of spherical harmonics. The Fourier components are defined as,
\begin{eqnarray}
T(\bi l)&=&\int d^2\bi \theta  e^{i\bi l \cdot \bi
\theta} T(\bi \theta) \nonumber \\
\delta \bi \theta(\bi l)&=&\int d^2\bi \theta  e^{i\bi
l \cdot \bi \theta} \delta \bi \theta(\bi \theta) \nonumber \\
\end{eqnarray}
We assume both the CMB and the deflection angle are Gaussian random
fields characterized by their power spectra,
\begin{eqnarray}
\langle \tl T({\bi l}_1) \tl T({\bi l}_2) \rangle &=& (2\pi)^2
\delta^D ({\bi l}_{12}) C_l^{\tl T \tl T} \nonumber \\ \langle
\delta \bi \theta({\bi l}_1) \cdot \delta \bi \theta({\bi
l}_2)\rangle &=& (2\pi)^2 \delta^D ({\bi l}_{12}) C_l^{\delta
\delta},
\end{eqnarray}
where $\bi l_{12}=\bi l_1 + \bi l_2$,
$C_l^{\tl T \tl T}$ is the power spectrum of the primary CMB
anisotropies and $C_l^{\delta \delta}$ is the power spectrum of the
deflection angle.

To leading order in
the deflection angle
the Fourier components of the lensed CMB field are,
\begin{eqnarray}
T({\bi l})&=&\tl T({\bi l}) + \int {d^2{\bi l}^\prime \over
(2\pi)^2}\ \delta {\bi \theta}({\bi l}^\prime)\cdot\nabla \tl T(
{\bi l}-{\bi l}^\prime). \label{convolution}
\end{eqnarray}
We want to calculate the power spectrum of the lensed CMB field
($C_l^{TT}$) on very small scales, scales much smaller than the
coherence length of the unlensed CMB gradient. We start by
considering a small patch of the sky of size $L$, with $L$  small
enough that the gradient of the unlensed field can be considered
constant so that the unlesed CMB map can be approximated linearly.
On these scales, the small scale variations of the deflection
angle generate additional power in the lensed CMB field. We define
$\tl T_x(\bi \theta)=\tl T_{x0}$ and $\tl T_y(\bi \theta)=\tl
T_{y0}$ so that,
\begin{eqnarray}
T({\bi \theta})&=&\tl T_{x0}\ \ 
[\theta_x + \delta \theta_x({\bi \theta})] +
\tl T_{y0}\ \  [\theta_y + \delta \theta_y({\bi \theta})].
\end{eqnarray}
For the Fourier modes at large $l$ we get,
\begin{eqnarray}
T({\bi l})&=&\tl T_{x0} \delta \theta_x({\bi l}) +\tl T_{y0} \delta
\theta_y({\bi l}).
\end{eqnarray}
Thus the power spectra of the lensed temperature is given by,
\begin{eqnarray}
\langle T({\bi l}_1) T({\bi l}_2) \rangle = (2\pi)^2 \delta^D
({\bi l}_{12}) \sigma_S C_{l_1}^{\delta \delta}/2,
\label{tailpower}
\end{eqnarray}
with $\sigma_S=\langle\tl T_{x0}^2+\tl T_{y0}^2\rangle$. For SCDM
$\sigma_S \approx 2\times 10^9 \mu K rad^{-1}\approx 13 \mu K
{\rm arcmin}^{-1}$. Note that to obtain equation (\ref{tailpower}) we
have replaced $\tl T_{x0}^2$ and $\tl T_{y0}^2$ by their averages.
This implies that we are measuring the power spectrum over a large
enough area of the sky that there are many patches over which the
gradient is averaged. Equation (\ref{tailpower}) is a very
interesting result on its own, it shows that the power in the tail
of the CMB anisotropies directly gives the power spectrum of the
deflection angle.  This result can also be derived by taking the
appropriate limit in the full expression for the lensed CMB
spectra as derived for example in \cite{uroslens}.

\begin{figure*}
\begin{center}
\leavevmode \epsfxsize=6.0in \epsfbox{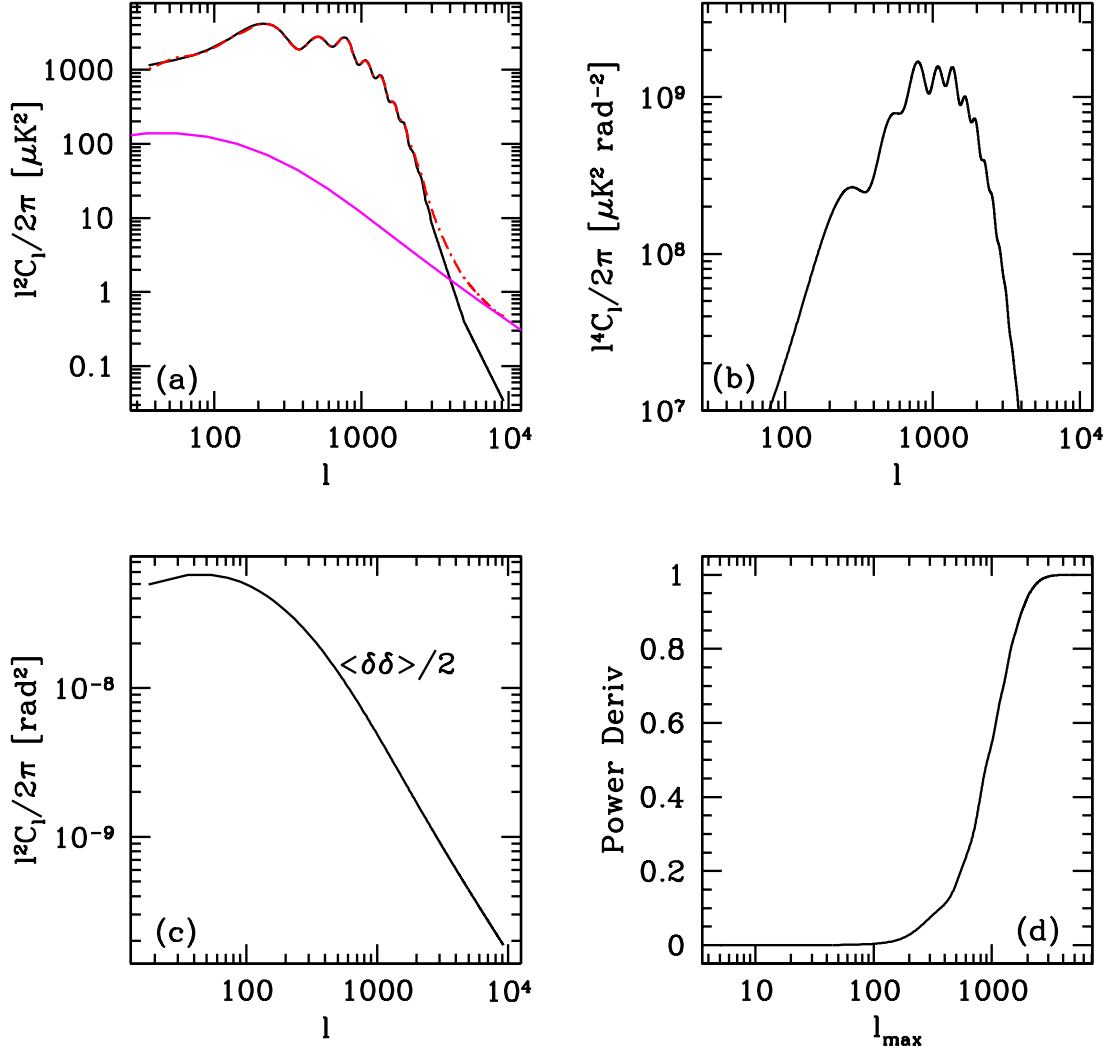}
\end{center}
\caption{The panel (a) shows the lensed and unlensed temperature
power spectra together with $\sigma_S C_l^{\delta \delta}/2$.
Panel (b) the power spectra of the derivative of the unlensed CMB
field, (c) the spectra of $\delta {\bi \theta}$ and (d) the
cumulative power in the CMB derivative.} \label{spectra}
\end{figure*}

In figure \ref{spectra} we show the various power spectra so as to
gain some insight on the orders of magnitude. In panel a we show
the lensed and unlesed CMB spectra together with $\sigma_S
C_l^{\delta \delta} /2$, the limiting value in the damping tail.
Several conditions must be met for equation (\ref{tailpower}) to
be valid. A very important fact is that the power spectra of both
the deflection angle and the CMB derivative fall with $l$. This
means that when we look at the power generated by lensing in a
high $l$ mode of the temperature, it could be dominated by  a
nearby $l$ of the temperature derivative multiplied by a low $l$
deflection angle mode rather than by the power coming from a small
$l$ temperature derivative mode multiplied by a high $l$
deflection mode. Equation (\ref{tailpower}) is only valid in the
latter case. We attain this limit because the power spectra of the
deflection angle falls more slowly with $l$ than that of the CMB
derivative. The power in the deflection angle falls like a power
law (figure \ref{spectra}c) while than in the CMB derivative drops
exponentially (figure \ref{spectra}b). As a consequence at a high
enough $l$ it is always easier to produce power by multiplying low
frequency temperature modes by high frequency deflection modes. At
$l>4000$ this effect dominates, at lower $l$ both effect compete.
Another condition is that we should apply equation
(\ref{tailpower}) for $l$ values large enough that most of the
power in the unlensed derivative (contributing to $\sigma_S$)
comes from a lower $l$. Figure \ref{spectra}d shows that for SCDM
most of the power in the derivative comes from scales $l<2000$.

\section{Correlation between small and large scales}\label{seccross}

In the previous section we have shown that the power in the CMB
anisotropies at large enough $l$ measures the power spectrum of
the deflection angle $\delta \bi \theta$. If we want to use the
small scale power to measure the fluctuations in the deflection
angle we need a way to determine that the power we observe is
indeed produced by lensing and is not at the last scattering
surface (ie. it is due to a different underlying primary
anisotropies spectrum) or that it is not produced by another
secondary effect. To be able to test this we need to consider
higher order statistics, we need to go beyond the power spectrum.

The crucial point we are going to use is that the power produced
by lensing comes from large scale primary temperature modes
promoted to higher $l$ by the small scale deflection of the
photons. We will attempt to test the hypothesis that the power we
observe is generated by lensing by looking at the correlation
between the large scale temperature derivative and the small scale
power. If we continue with the simplification that the gradient of
the unlensed field is constant, when we high pass filter the
temperature map we obtain,
\begin{eqnarray}
T_H(\bi \theta)&=& \int_{l_{h1}}^{l_{h2}} {d^2\bi l \over
(2\pi)^2} e^{-i\bi l \cdot \bi \theta} T(\bi l) \nonumber \\
&\approx&\tl T_x \delta\theta_x(\bi \theta) + \tl T_y
\delta\theta_y (\bi \theta),
\end{eqnarray}
where $l_{h1}$ and $l_{h2}$ define a ring in $l$ space and
we have abused notation and called $\delta\theta_i(\bi \theta)$ the
high passed filtered deflection angle.
We will study the behavior of the square of $T_H$ properly smoothed,
\begin{eqnarray}
{\cal H}({\bi \theta})&=& \int d^2{\bi \theta}^{\prime}
W_H({\bi \theta}-{\bi \theta}^{\prime}) T^2_H({\bi
\theta}^{\prime}).
\end{eqnarray}
If we smooth the map over a scale bigger than that of the
variations in the deflection angle but over which $\tl T_x$ and
$\tl T_y$ remain constant, we can replace $\delta\theta_x^2$ and
$\delta\theta_y^2$ by their averages. It follows that the smoothed
high passed filtered map is approximately,
\begin{eqnarray}
{\cal H}(\bi \theta)&\approx& {\langle
\delta\theta^2 \rangle\over 2}(\tl T_x^2 + \tl T_y^2)(\bi \theta).
\label{hmap}
\end{eqnarray}
Only the modes with wavevectors $l_{h1}<l<l_{h2}$ contribute to the
average $\langle
\delta\theta^2 \rangle$ in equation (\ref{hmap}).

We now look at the anisotropies on larger scales by applying a low
pass filter to the temperature field,
\begin{eqnarray}
T_L(\bi \theta)&=& \int^{l_{l}}_0 {d^2\bi l \over (2\pi)^2}
e^{-i\bi l \cdot \bi \theta} T(\bi l).
\end{eqnarray}
We compute
the square of the  gradient,
\begin{eqnarray}
{\cal L}(\bi \theta)&=&||\nabla T_L||^2 \nonumber \\ &\approx&
(\tl T_x^2 + \tl T_y^2)(\bi \theta). \label{lmap}
\end{eqnarray}
Equations (\ref{hmap}) and (\ref{lmap}) show
the maps of ${\cal L}$ and ${\cal H}$ trace each other,
\begin{eqnarray}
{\cal H}(\bi \theta)\approx {\langle \delta\theta^2 \rangle\over
2} {\cal L}(\bi \theta). \label{lhrelation}
\end{eqnarray}

We have a very simple test to establish that the small scale power
is generated by lensing, the small scale power has to be
correlated with the large scale derivative. Instead if the small
scale power is Gaussian power already present at the last
scattering surface, different Fourier modes are uncorrelated. In
the Gaussian case if we construct ${\cal L}$ and ${\cal H}$ with
different modes ($l_{h1}> l_l$) there will be no correlation
between the two maps.  In practice
instead of smoothing the high passed temperature map we work in
Fourier space and correlate the low $l$ modes of $T_H^2(\bi
\theta)$ with those of the ${\cal L}(\bi \theta)$.

Equations ({\ref{hmap}) and ({\ref{lmap}) were obtained under
simplifying assumptions, more general expressions can also be
derived but we only show this limiting case to make the physics
more transparent. Comparison with simulations show that this
expressions are good approximations in the regime of interest. The
procedure to simulate a lensed CMB map was discussed in detail in
reference \cite{longlens}. We generate  realizations of CMB and
projected mass density $\kappa$ and use a ray tracing technic to
produce a lensed CMB map.

We will  explain the effect we are studying by first presenting an
toy example in 1 dimension. We will consider a 2 dimensional
universe so that the last scattering surface becomes a line. In
the top panel of figure \ref{1dfig} we show the unlensed CMB field
together with the low passed filtered ($l_l=2000$) lensed
anisotropies. It is clear that both curves trace each other,
although they are displaced. This is the consequence of the large
scale modes of the deflection angle, but remember we are after the
power generated by the small scale modes of $\delta\bi \theta$. In
the middle panel we show the high passed ($l>6000$) lensed CMB,
the fluctuations in this panel are the result of lensing. The
power is not distributed uniformly, the fluctuations are larger
where the derivative of the CMB in the top panel is larger, as equation
(\ref{hmap}) indicates. This is the same physical effect we
discussed in the cluster example. On the contrary in regions where
the CMB is constant, surface brightness conservation implies that
lensing does not create any power. The bottom panel shows the
square of the small scale temperature anisotropies and a scaled
version of the low $l$ gradient to ease the comparison.

\begin{figure*}
\begin{center}
\leavevmode \epsfxsize=6.0in \epsfbox{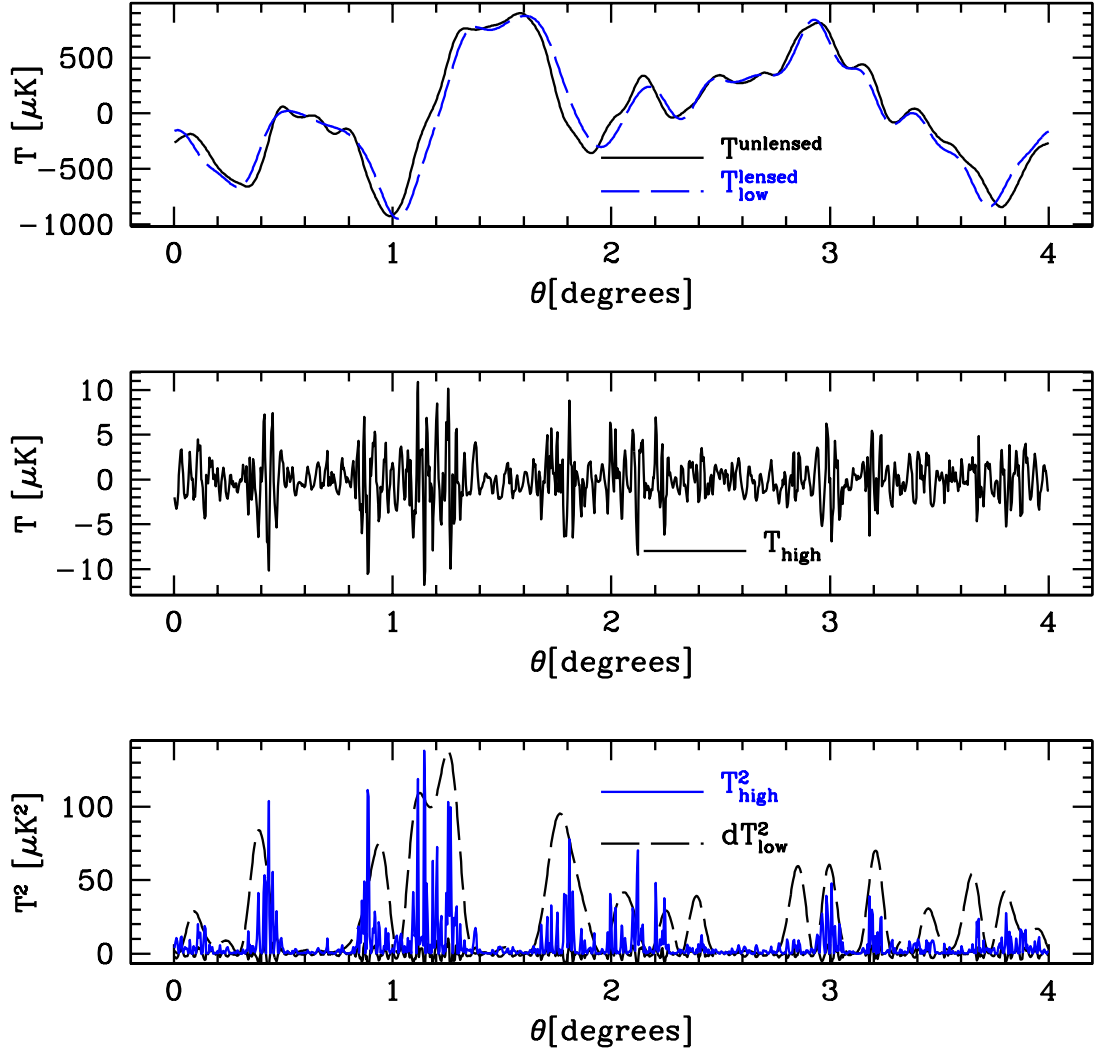}
\end{center}
\caption{Example of the generation of power by gravitational
lensing in one dimension. The upper panel shows the unlensed
temperature and the result of filtering the lensed $T$. The middle
panel shows the high $l$ power generated by lensing. In the bottom
we show the square of the small scale power and the square of the
large scale gradient (arbitrarily scaled). } \label{1dfig}
\end{figure*}

We can now look at the results for a 2 dimensional last scattering
surface shown in figure \ref{maps}. The low passed gradient map
was constructed using scales $l<2000$ and the high passed
temperature was constructed with modes $l>6000$. The upper left
panel shows the lensed CMB map while the upper right panel has the
high passed filtered field. It is clear that the small scale power
in not Gaussian and is higher where the large scale derivative is
higher. To make this even more apparent the two bottom panels show
the $\cal L$ and $\cal H$ maps, the correlation is excellent.

\begin{figure*}
\begin{center}
\leavevmode \epsfxsize=6.0in \epsfbox{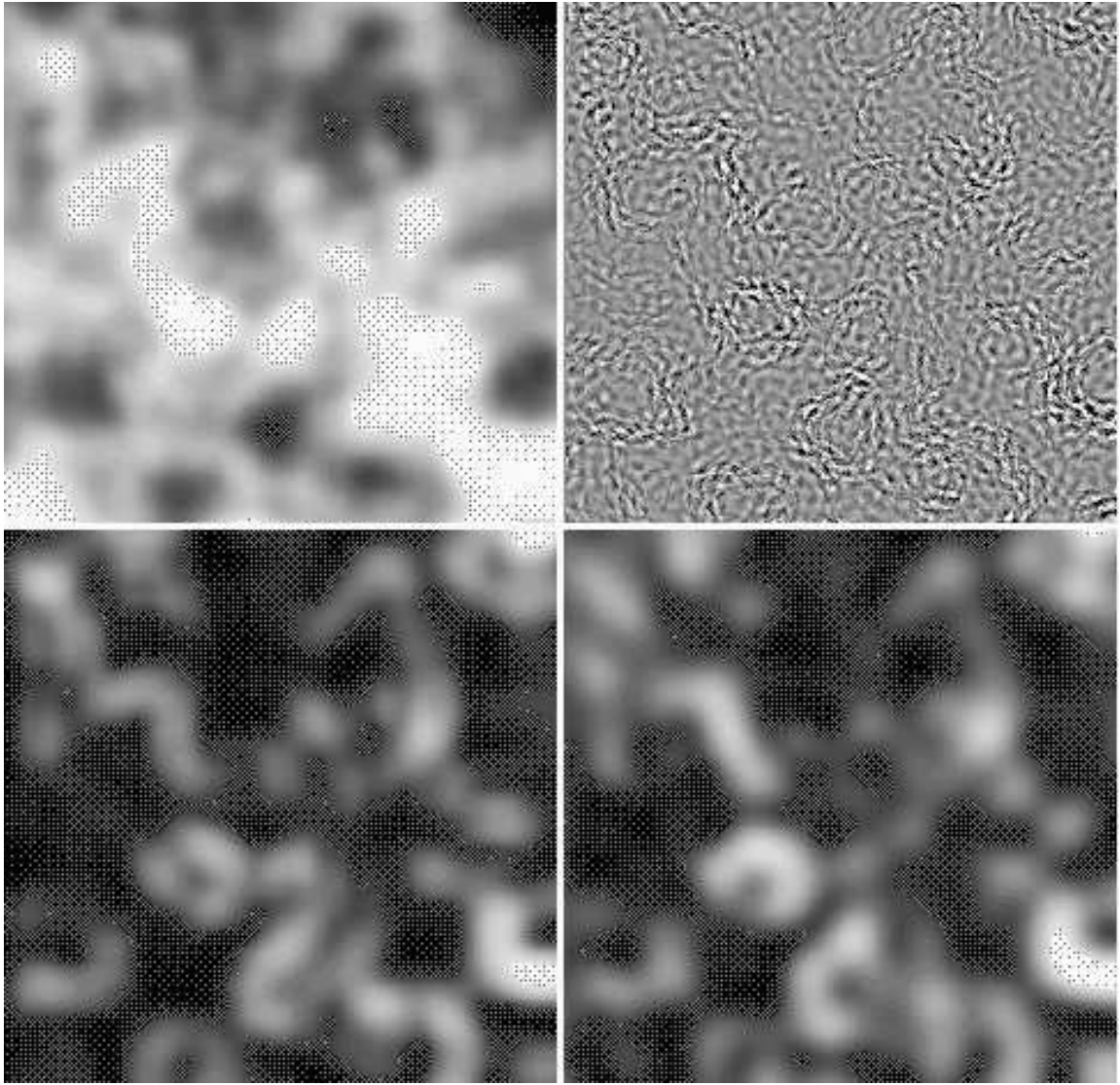}
\end{center}
\caption{The upper panel on the left
shows the  lensed temperature field.
The upper right panel shows the high passed temperature.
Bottom left has gradient square and bottom right
smoothed square of high passed temperature.}
\label{maps}
\end{figure*}

The $\cal L$ and $\cal H$ maps only differ by a factor $\langle
\delta \theta^2\rangle/2$ (equation \ref{lhrelation}), so their
cross correlation defined as $\langle {\cal H} (\bi l_1){\cal L}
(\bi l_2)\rangle=(2\pi)^2 \delta^D(\bi l_{12}) C_{l_1}^{\cal HL}$,
is simply related to the power spectrum of the $\cal L$ map
($C_l^{\cal LL}$),
\begin{eqnarray}
C_l^{\cal HL}&\approx&{\langle \delta \theta^2\rangle \over 2}
C_l^{\cal LL} \nonumber \\ C_l^{\cal HH}&\approx&{\langle \delta
\theta^2\rangle^2 \over 4} C_l^{\cal LL} \label{relcross}
\end{eqnarray}
In figure \ref{corr}a we show the ratio $C_l^{\cal HL}/C_l^{\cal
 LL}$ obtained in simulations. We have used $l<2000$ to construct
 $\cal L$, and several
 different rings at higher $l$ for $\cal H$.  The order of magnitude of
the ratio is consistent with equation (\ref{relcross}), it remains
constant over a wide range of $l$ but starts to fall once we approach
 $l\sim 2000$. The first relation in equation (\ref{relcross}) is really
 $C_l^{\cal HL}=W_l {\langle \delta \theta^2
 \rangle} C_l^{\cal LL}/2$ where $W_l$ is some window function. The origin
of the window can be understood as follows: lets call ${\bi l}_h$
a high $l$ mode of the deflection and ${\bi l}_1$ and ${\bi l}_2$
two modes of the derivative. When we compute the gradient square
this modes combine to give power at ${\bi l}_3={\bi l}_1+{\bi
l}_2$. In the high passed case, we recover this component of the
field by looking at the modes with ${\bi l}_h+{\bi l}_1$ and
$-{\bi l}_h+{\bi l}_2$ so that when we multiply them they give the
${\bi l}_3$ variations. As ${\bi l}_3$ becomes larger at least one
of ${\bi l}_1$ or ${\bi l}_2$ also become  larger and
 because we are selecting only a
ring in $l$ space for the high passed map there are higher chances
either ${\bi l}_h+{\bi l}_1$ or $-{\bi l}_h+{\bi l}_2$ will fall
 outside the ring and we cannot reconstruct ${\bi l}_3$. So as
${\bi l}_3$ gets larger  the correlation between the two maps fall.
This effect explains the window $W_l$ is less important the
wider rings for the high passed maps is (compare the high $l$
 behavior of the $4000-12000$ and $4000-8000$
rings in figure \ref{corr}a).

In figure \ref{corr}b we show the correlation coefficient ${\cal
C}_l\equiv C_l^{\cal H L}/ \sqrt{(C_l^{\cal H H} C_l^{\cal L
L})}$. Note that the cross correlation for low $l$ is very high,
almost one. This proves our claim that on large scales the $\cal
L$ and $\cal H$ maps trace each other almost perfectly. There are
several reasons why the two maps do not correlate exactly. First,
although most of the power in the derivative of the CMB comes from
$l<2000$ there is still about $5\%$ additional power coming from
higher $l$. This degrades the cross-correlation because some of
the power on the high passed map is coming from these modes of the
derivative field. A more extreme case is shown in the figure,
where only modes with $l<1000$ were used to construct the $\cal L$
map, the cross-correlation in this case is significantly smaller.
The other important effect is that some of the power in the high
$l$ map is due to primary anisotropies, these modes are
uncorrelated with the low $l$ derivative and thus reduce the cross
correlation coefficient. This effect  is more important for the
$4000-8000$ ring than for the $8000-12000$ so the cross
correlation is smaller for the former. The $4000-12000$ rings
falls somewhere in the middle of this two cases.

\begin{figure*}
\begin{center}
\leavevmode \epsfxsize=6.0in \epsfbox{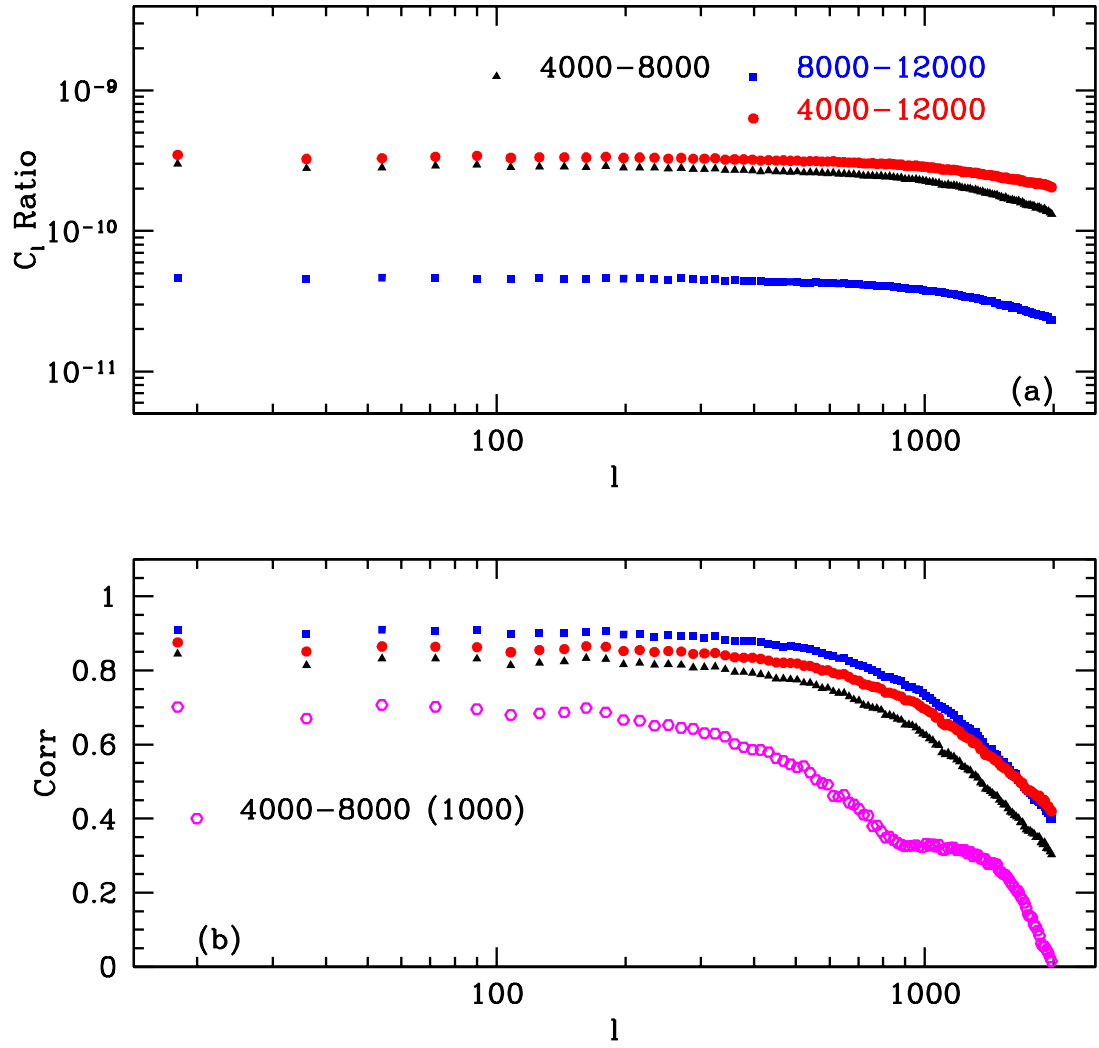}
\end{center}
\caption{Panel (a) shows the ratio $C_l^{\cal H L}/C_l^{LL}$ for
several different rings used to construct $\cal H$. Panel (b)
shows the cross correlation coefficient in $l$ space.}
\label{corr}
\end{figure*}

We will calculate $C_l^{\cal LL}$, which can be done analytically.
If we assume that only a Gaussian component is contributing to the
low pass filtered field then it is possible to calculate the power
spectra of the gradient squared, ${\cal L }=|| \nabla T_L ||^2$.
The correlation function in real space for two points separated an
angle $\theta$ in the $\bi x$ direction gives,
\begin{eqnarray}
\langle {\cal L}(0){\cal L}(\theta)\rangle &=& \langle {\cal L}^2
\rangle^2+ 2 (C_{xx}^2(\theta)+C_{yy}^2(\theta)). \label{lcorr}
\end{eqnarray}
We have introduced $C_{xx}(\theta)=\langle T_x(0)T_x(\theta
)\rangle$ and $C_{yy}(\theta)=\langle T_y(0)T_y(\theta )\rangle$.
They are given by,
\begin{eqnarray}
C_{xx}(\theta)&\equiv&\langle \tl T_x(0) \tl T_x(\theta)
\rangle_{CMB} \nonumber \\ &=&(2\pi)^{-2}\int^{l_l} d^2{\bi l}\
e^{i l\cdot \theta \cos\phi_l}\ l^2C^{\tl T \tl T}_l \cos^2\phi_l
\nonumber
\\
&=&\int_0^{l_l} {l dl \over 4 \pi}\ l^2C^{\tl T \tl T}_l \left[
J_0(l\theta)-J_2(l\theta)\right]\nonumber \\ &\equiv&{1 \over
2}[C_0(\theta)-C_2(\theta)] \nonumber \\
C_{yy}(\theta)&\equiv&\langle \tl T_y(0) \tl T_y(\theta)
\rangle_{CMB} \nonumber \\ &=&(2\pi)^{-2}\int^{l_l} d^2{\bi l}\
e^{i l\cdot \theta \cos\phi_l}\ l^2C^{\tl T \tl T}_l \sin^2\phi_l
\nonumber
\\
&=&\int_0^{l_l} {l dl \over 4 \pi}\ l^2C^{\tl T \tl T}_l \left[
J_0(l\theta)+J_2(l\theta)\right] \nonumber \\ &\equiv&{1 \over
2}[C_0(\theta)+C_2(\theta)] \nonumber \\
C_{xy}(\theta)&\equiv&\langle \tl T_x(0) \tl T_y(\theta)
\rangle_{CMB} \nonumber \\ &=&(2\pi)^{-2}\int^{l_l} d^2{\bi l}\
e^{i l\cdot \theta \cos\phi_l}\ l^2C^{\tl T \tl T}_l \cos\phi_l
\sin\phi_l \nonumber \\ &=& 0, \label{correlations}
\end{eqnarray}
where $C_0(\theta)$ and $C_2(\theta)$ are defined as the integrals
over $l^3C_l^{\tl T\tl T}dl/2\pi$ weighted with $J_0(l \theta)$
and $J_2(l\theta)$, respectively.

The constant part in equation (\ref{lcorr}) only contributes to
the $l=0$ mode. The second term is $\sigma_S^2 N^{\cal S \cal
S}(\theta)$ in the notation of \cite{longlens}. The power spectra
of the low pass filtered map is that of $\cal S$ in
\cite{longlens}, but with only the low $l$ modes included. For
$l\neq 0$,
\begin{eqnarray}
C_l^{\cal L L}&=&2\pi \int \theta d\theta (C_o^2+C_2^2)
J_0(l\theta). \label{llfinal}
\end{eqnarray}
This calculation coincides perfectly with the results of
simulations. In the presence of detector noise the above
expression generalizes to $C_l^{\cal L L}\rightarrow C_l^{\cal L
L}+N_l^{\cal L L}$, where $N_l^{\cal L L}$ is the detector noise
contribution calculated using equations (\ref{correlations}) and
(\ref{llfinal}) with the power spectrum of the detector noise.

Let us now consider $C_l^{\cal H H}$. If the contribution to the power
where Gaussian then the correlation function of $\cal H$ in real space
would be,
\begin{eqnarray}
\langle {\cal H}(0){\cal H}(\theta)\rangle &=& \langle {\cal H}^2
\rangle^2+ 2 C^2(\theta) \label{hcorr}
\end{eqnarray}
where,
\begin{eqnarray}\label{hcorr2}
C(\theta)
&\equiv&\langle \tl T_H(0) \tl T_H(\theta) \rangle_{CMB} \nonumber \\
&=&(2\pi)^{-2}\int d^2{\bi l}\
e^{i l\cdot \theta \cos\phi_l}\ C^{\tl T \tl T}_l  \nonumber \\
&=&\int_{l_{h1}}^{l_{h2}}\ {l dl \over 2 \pi}
C^{\tl T \tl T}_l  J_0(l\theta).
\end{eqnarray}
The power spectrum becomes,
\begin{eqnarray}
C_l^{\cal H H}&=&4\pi \int \theta d\theta C^2(\theta) J_0(l\theta)
\label{hhfinal}
\end{eqnarray}
This expression only coincides with the results of simulations
when either detector noise or intrinsic CMB anisotropies dominate
the power in the $l$ range used to construct $\cal H$, that is if
the unlensed power is Gaussian. In the idealized detector noise
free examples we have discussed above this corresponds to $l_{h2}
< 4000$. Fortunately for the power generated by lensing, the power
spectra of the $\cal H$ map is a scaled version of the power
spectrum of $\cal L$, thus it can be obtained using equations
(\ref{relcross}) and (\ref{hcorr2}). Again in the presence of
noise $C_l^{\cal H H}\rightarrow C_l^{\cal H H}+N_l^{\cal H H}$
with $N_l^{\cal H H}$ calculated using equation (\ref{hhfinal})
with the detector noise power spectra.

To assess the signal to noise of our lensing signal we need to
calculate the variance in the estimator of $C_l^{\cal HL}$,
\begin{eqnarray}\label{hlestim}
\hat C_l^{\cal HL}={A_f\over (2\pi)^2}\int {dA_l\over A_l} {\cal
H}(\bi l){\cal L}(-\bi l),
\end{eqnarray}
where the integral is done over a small area in $l$ space of size
$A_l$ centered around $l$ and $A_f=(2\pi)^2/\Omega$ is the area of
the fundamental cell in $l$ space. We have denoted $\Omega$ the
area of sky observed.

If we assume that $\cal L$ and $\cal H$ are Gaussian fields, we
can calculate the variance as,
\begin{eqnarray}
{\rm Cov}[(\hat C^{\cal H L}_l)^2] &=& {A_f\over A} [(C^{\cal H
L}_l)^2+C^{\cal H H}_l C^{\cal L L}_l]. \label{sigma}
\end{eqnarray}
The ratio ${A/A_f}$ is the number of available modes one can use
to measure the cross correlation. It can be approximated by
$A_l/A_f=f_{sky}(2l+1)$ for a unit width ring. In figure
\ref{variances} we show the variance in the cross correlation
measured in simulations normalized to the Gaussian prediction in
equation (\ref{sigma}). The agreement is very good implying that
we can use the Gaussian formula to compute the variance.
\begin{figure*}
\begin{center}
\leavevmode \epsfxsize=4.0in \epsfbox{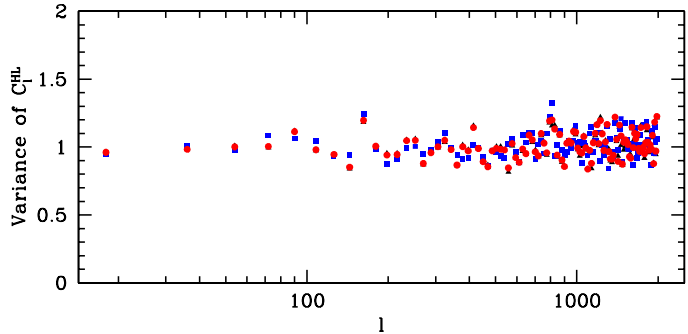}
\end{center}
\caption{Ratio of variance of power spectrum measured in the
simulation to the predicted result assuming Gaussian fields.}
\label{variances}
\end{figure*}

To compute the total S/N we want to combine the signal in all the
$l$  modes  of the cross correlation that we can measure. To do
this we will compute $\hat X=\sum_l \alpha_l \hat C^{\cal HL}_l$
choosing $\alpha_l$ to maximize S/N$=[\langle\hat X \rangle
^2/{\rm Cov}(\hat X^2) ]^{1/2}$. It is straight forward to show
that $\alpha_l= C^{\cal H L}_l / {\rm Cov}[(C^{\cal H L}_l)^2]$,
\cite{iswletter}. We then get,
\begin{eqnarray}\label{s2n}
{ S \over N} &=& \left[\sum_{l<l_l} {(C^{\cal H L}_l)^2 \over {\rm
Cov}[(C^{\cal H L}_l)^2]} \right]^{1/2} \nonumber \\ &=& \left[
f_{sky} \sum_{l<l_l} (2l+1) {{\cal C}_l^2 \over ((1+N_l^{\cal
LL}/C_l^{\cal LL})(1+N_l^{\cal HH}/C_l^{\cal HH})+{\cal C}_l^2)}
\right]^{1/2}.
\end{eqnarray}
We have implicitly assumed rings of unit width.

We want to relate equation (\ref{s2n}) to the S/N with which we
can measure the small scale power generated by lensing. It is
straightforward to show using equations (\ref{llfinal}),
(\ref{hhfinal}) and the fact that ${\cal C}_l\sim 1$  that
$N_l^{\cal HH}/C_l^{\cal HH}\approx (w_T^{-1}/\bar C^{TT})^2$,
here $\bar C^{TT}$ is the averaged power in the band
$l_{h1}<l<l_{h2}$ and $w_T^{-1}$ is the power spectrum of the
detector noise, assumed to be white noise.

We will consider the limit in which $w_T^{-1}$ is small enough
that $N_l^{\cal LL}\ll C_l^{\cal LL}$ but is large enough that
$N_l^{\cal HH}\gg C_l^{\cal HH}$. If the noise where even lower
then the S/N for detecting the cross correlation will be large
S/N$\sim N_L^{1/2}$ where $N_L$ is the total number of cross
correlations that can be measures, $N_L\approx f_{sky}l_l^2$. In
the limit we are considering,
\begin{eqnarray}
({S \over N})^2 &=& N_L {\cal C}_l^2 ({\bar C_l \over w_T^{-1}})^2 \nonumber
\\
    &=& 2 {N_L \over N_P}{\cal C}_l^2 ({S \over N})^2_P,
\end{eqnarray}
where we have introduced  $({S \over N})_P^2=N_P/2 (\bar C_l /
w_T^{-1})^2$, the signal to noise with which we can measure the
power in the band $l_{h1}-l_{h2}$. We call $N_P$ the number of
modes that we can use to estimate the power, which is relate to
$N_L$ by,
\begin{equation}
{N_L\over N_P}={l_l^2 \over l_{h2}^2 - l_{h1}^2}.
\end{equation}
Thus the S/N to measure the cross correlation is comparable
although always somewhat smaller than that to detect the power
directly. For example if we take $l_{l}=2000$, $l_{h1}=4000$ and
$l_{h2}=6000$ then $S/N\approx \sqrt{2/5}{\bar{\cal C}_l} (S/N)_P$.

\section{The three and four point functions in the small angle limit}

In this paper and in our previous studies 
\cite{psdmletter,longlens,iswletter} we have
investigated several ways of detecting the effect of gravitational
lensing on the CMB. As we have explained this amounts to trying to
detect the distortions on the random CMB maps 
created by the random distributions of the dark
matter in the universe. In \cite{iswletter} we used the ISW effect as
a tracer of the dark matter distribution and combinations of the CMB
derivatives to measure the effect of lensing. The cross correlation of
these two effects allowed us to gain information about the time
evolution of the gravitational potential. Our method combined the
information in 
particular configurations of the three point function of
the temperature. Other studies have used other combinations of the
bispectrum to detect the signal \cite{goldsper} and also calculated
the contributions to the bispectrum coming from other secondary
procesess \cite{goldsper,wayne}. 

In \cite{longlens} we used the
power spectrum of quadratic combination of derivatives of the CMB to
measure the power spectrum of the projected mass density
$\kappa$. This method was valid in the limit in which we wanted to
recover the long wavelength modes of $\kappa$ from information in the
small scale CMB. This regime is analogous to weak lensing of
background galaxies. In essence the different estimates of the 
power spectrum of $\kappa$ at different scales were obtained by
combining different configurations of the four point function of the
lensed temperature.

In the present paper we studied other configurations of the four point
function to illustrate the nature of the non-Gaussianities induced by
lensing on small scales. The non-Gaussian nature of the generated
power manifested itself in the correlations between the large scale
gradient and the small scale generated power.  

In order to have a unified picture of the different statistics we have
proposed it is convenient study directly the four point function of
the temperature field and a three point function which correlates two
temperatures and another field $X$. The field $X$ stands for any
field that cross correlates with $\kappa$. In our paper
\cite{iswletter} $X=T$ but one can imagine doing this correlation with
other tracers of the mass, like the fluctuations of the Far
Infrared Background \cite{knoxhaimanzal}.

We define the connected three and four point functions as, 
\begin{eqnarray}
\langle X(\bi l_1) T(\bi l_2)T(\bi l_3)\rangle_c&=&(2\pi)^2 \delta^D(\bi
l_{123}) T_3(\bi l_1,\bi l_2,\bi l_3) \nonumber \\
\langle T({\bi l}_1) T({\bi l}_2) T({\bi l}_3) T({\bi l}_4)
\rangle_c &=& (2\pi)^2\delta^D ({\bi l}_{1234}) T_4(\bi l_1,\bi l_2,\bi
l_3,\bi l_4), 
\label{threefourpointdef}
\end{eqnarray}

Gravitational lensing produces, 
\begin{eqnarray}
T_3(\bi l_1,\bi l_2,\bi l_3)
&=&2C_{l1}^{\kappa X} [C^{\tl T \tl T}_{l2} {\bi l_2 \cdot \bi l_1
\over l_1^2} +C^{\tl T \tl T}_{l3} {\bi l_3 \cdot \bi l_1
\over l_1^2} ] \nonumber \\
T_4({\bi l}_1,{\bi l}_2,{\bi l}_3,{\bi
l}_4)&=&C^{\tl T \tl T}_{l_1} C^{\tl T \tl T}_{l_2}
\left[{(\bi l_1 + \bi l_3)\cdot \bi l_1 (\bi l_1 + \bi l_3)\cdot
\bi l_2 \over ||\bi l_1 + \bi l_3||^2}C^{\delta \delta}_{l_{13}} +
 {(\bi l_1 + \bi l_4)\cdot \bi l_1 (\bi l_1 + \bi l_4)\cdot \bi
l_2 \over ||\bi l_1 + \bi l_4||^2}C^{\delta\delta}_{l_{14}}
\right] \nonumber \\ &+& {\rm permutations}\ (5\ {\rm terms\
proportional\ to}\ C^{\tl T\tl T}_{l_1} C^{\tl T \tl
T}_{l_3},C^{\tl T \tl T}_{l_1} C^{\tl T \tl T}_{l_4},C^{\tl T \tl
T}_{l_2} C^{\tl T \tl T}_{l_3},C^{\tl T \tl T}_{l_2} C^{\tl T \tl
T}_{l_4},C^{\tl T \tl T}_{l_3} C^{\tl T \tl T}_{l_4}).
\label{fullthreefour}
\end{eqnarray}

The unconnected part of the four point function also gets corrections.
To make the calculation of these terms  fully
consistent up to second order in the deflection angle we need to also 
consider the contributions coming from the second order in the
expansion of equation (\ref{expansion}). The unconnected terms are not
relevant for our study so we will not write them down here.


In our previous papers we introduced three variables $\cal E$, $\cal
B$ and $\cal S$. We had defined them in terms of derivatives to the
temperature field. Equivalently we can write,  
\begin{eqnarray}
{\cal S}(\bi l)&=&\int {d^2\bi l_1\over (2\pi)^2} (\bi l - \bi l_1)\cdot
\bi l_1 T(\bi l - \bi l_1) T(\bi l_1) \nonumber \\
{\cal Q}(\bi l)&=&\int {d^2\bi l_1\over (2\pi)^2} [(l_x - l_{1x})
l_{1x}-(l_y - l_{1y})
l_{1y}] T(\bi l - \bi l_1) T(\bi l_1) \nonumber \\
{\cal U}(\bi l)&=&\int {d^2\bi l_1\over (2\pi)^2} [(l_x - l_{1x})
l_{1y}+(l_y - l_{1y})
l_{1x}] T(\bi l - \bi l_1) T(\bi l_1) \nonumber \\
&& \nonumber \\
{\cal E}(\bi l)&=&{\cal Q}(\bi l) \cos(2\phi_{\bi l})+{\cal U}(\bi l)
\sin(2\phi_{\bi l})\nonumber \\
&& \nonumber \\ 
{\cal B}(\bi l)&=&-{\cal Q}(\bi l) \sin(2\phi_{\bi l})+{\cal U}(\bi l)
\cos(2\phi_{\bi l})
\end{eqnarray}
When we average over the CMB random field we  get,
\begin{eqnarray}
\langle {\cal S} (\bi l) \rangle_{CMB} &=&((2\pi)^2 \delta^D(\bi l) -2
\kappa(\bi l))\sigma_{\cal S} \nonumber \\
\langle {\cal E} (\bi l) \rangle_{CMB} &=&-2
\kappa(\bi l) \sigma_{\cal S} \nonumber \\
\langle {\cal B} (\bi l) \rangle_{CMB} &=& 0.
\end{eqnarray}
We have introduced $\sigma_{\cal S}=\int l dl/2\pi \ l^2C_l^{\tl T \tl
T} $.

To extract all the information in this three point function se combine
all possible configurations with a weight $\beta$ chosen to maximize
the signal to noise ratio.
We define, 
\begin{eqnarray}
\hat Y&=&{A_f\over (2\pi)^2}
\int {d^2\bi l_1\over A_l}{d^2\bi l_2\over A_l}
\beta(\bi l_1,\bi l_2,\bi l_3)
\ X(\bi l_1) T(\bi l_2)T(\bi l_3). \label{defz}
\end{eqnarray}
For these mildly non-Gaussian maps, 
the variance can be calculated by only taking the Gaussian part of the
temperature, so that
\begin{eqnarray}
{\rm Var}(\hat Y) &=&A_f  (2\pi)^2
\int {d^2\bi l_1\over A_l^2}{d^2\bi l_2\over A_l^2}
\beta^2({\cal S})\ 2 \ C^{X X}_{l_1} C^{\tl T \tl T}_{l_2} 
C^{\tl T \tl T}_{l_3}. 
\label{vary}
\end{eqnarray}
The power spectra in equation (\ref{vary}) must include the
contribution from detector noise. There are additional terms in the
variance if the field $X$ and $T$ had some cross correlation 
before lensing. In practice these terms are unimportant if one is
interested in measuring the cross correlation $C_l^{\kappa X}$ at
large angular scales (low $l$), as was the case in our study in
\cite{iswletter}. This is so because most the information of lensing
is encoded in the high $l$ modes of the temperature, so it is
effectively as if the integral over $\bi l_2$ in equation (\ref{defz})
is done over high $l$ modes while the $\bi l_1$ integral only involves
low $l$. Thus the terms that would involve $C_l^{TX}$ are absent
because there are no pair of triangles in which $X$ and $T$ are
evaluated on the same $\bi l$.

The weight $\beta$ that  maximizes the $S/N$ is $\beta\propto 
T_3(\bi l_1,\bi
l_2,\bi l_3)/2 C^{XX}_{l1} C^{\tl T \tl T}_{l_2}C^{\tl T \tl
T}_{l_3}$. Finally we get,
\begin{eqnarray}
({ S \over N})^2&=& A_f^{-1} \int d^2\bi l_1\int  { d^2\bi l_2 \over
(2\pi)^2} {T_3^2 \over 2 C^{X X}_{l_1} C^{\tl T \tl T}_{l_2} C^{\tl T
\tl T}_{l_3}} \nonumber \\
&=&  A_f^{-1} \int d^2\bi l_1 {4 (C_l^{\kappa X})^2 \over C^{X X}_l
C^{eff}_l} \nonumber \\
{1 \over C^{eff}_l }&\equiv& \int {d^2\bi l_2 \over (2\pi)^2} 
[C^{\tl T \tl T}_{l2} {\bi l_2 \cdot \bi l_1 \over l_1^2} +C^{\tl T
\tl T}_{l2} {\bi l_3 \cdot \bi l_1 \over l_1^2} ]^2 {1 \over 2 C^{\tl T
\tl T}(l_2) C^{\tl T \tl T}(l_3)}. 
\label{s2n3ptf}
\end{eqnarray}
The power spectra in the denominator include the contribution from
detector noise (amplified by the beam response $C_l\rightarrow C_l+B^2
N_l$. The easiest way to calculate $C^{eff}_l$ is to use a Monte Carlo
technique. We used the implementation of the VEGAS algorithm in
Numerical Recipes \cite{pressetal}.

The above formula can be compared to the what we obtain using the
$\cal S$ and $\cal E$ variables
\cite{iswletter}. Equation (7) of \cite{iswletter} read, 
\begin{equation}
({ S \over N})^2= \int d^2\bi l_1 {4 (C_l^{\kappa X})^2 \over C^{X X}_l}
W^2(l) ({1 \over N_l^{\cal S \cal S}}+{1 \over N_l^{\cal E \cal E}}),
\end{equation}
where $W^2(l)$ is a window that encapsulates the effect of beam
smearing. 
For low $l$ $N_l^{\cal S \cal S}$ and $N_l^{\cal E \cal E}$ are
constant and satisfy, $N_l^{\cal S \cal S}=2N_l^{\cal E \cal E}$.

\begin{figure*}
\begin{center}
\leavevmode \epsfxsize=6.0in \epsfbox{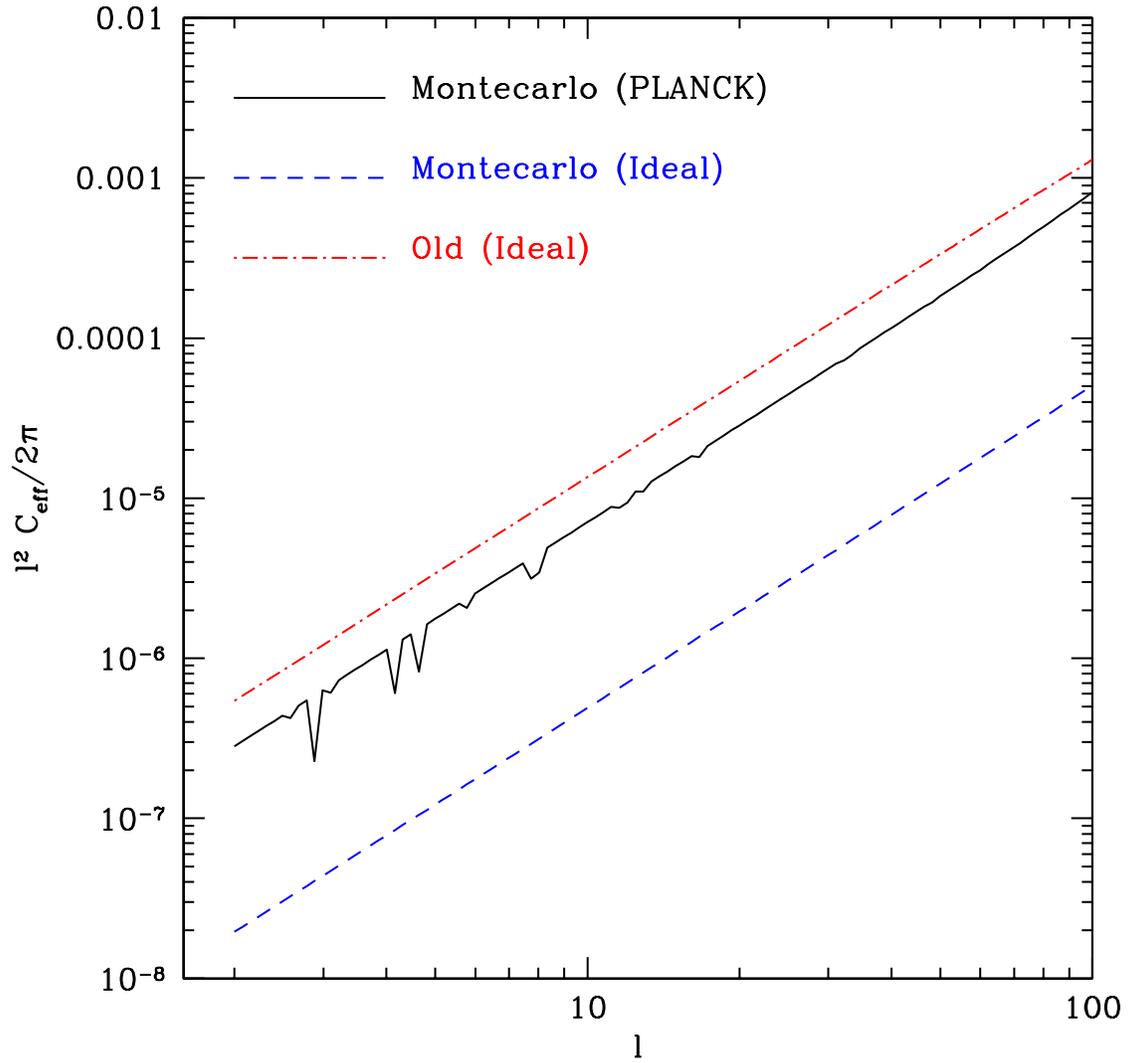}
\end{center}
\caption{Comparison between $C_l^{eff}$ and the noise of our old
method. We show the results for the Planck satellite and an ideal
experiment. For our old method there was no difference in the noise
between Planck and an ideal experiment on these angular scales.} 
\label{comp3pt}
\end{figure*}

In figure 
\ref{comp3pt} we compare $1/C^{eff}_l$ to $W^2(l) ({1 \over
N_l^{\cal S \cal S}}+{1 \over N_l^{\cal E \cal E}})$. We show the
results for two separate examples, an ideal experiment with no noise
and infinite resolution and the Planck satellite. We focus on the
large scale limit and for the ideal experiment we only consider the
information coming from modes with $l<3000$.
There are several salient features of the 
comparison. Although the difference
between the methods is not so large for Planck it is much larger for
the ideal experiment. This can be easily understood. In our previous
method the power spectrum of the CMB noise in this limit was,
\begin{equation}
N_l^{\cal S \cal S}=(2\pi){\int l^5 dl \ (C_l^{\tl T\tl T})^2 \over
(\int l^3 dl \ C_l^{\tl T\tl T})^2}.
\label{old1}
\end{equation}
It is clear from equation (\ref{old1}) that once we get into the
damping tail where the $C_l^{\tl T \tl T}$ fall exponentially, 
$N_l^{\cal S \cal S}$ no longer changes which means
that our method does not receive any information from those modes. In
contrast equation (\ref{s2n3ptf}) shows that the amplitude of 
the $C_l^{\tl T\tl T}$ cancels in $C^{eff}_l$ as long as the modes had
been measured with high $S/N$. Thus the new method
continues to extract information from modes in the damping tail. In
most practical cases this is not very important because the detector
noise quickly dominates in this regime and then all methods downweight
the modes. This explains why 
the difference between the two methods is not that
large for Planck. There is another relevant consideration. When
computing the variance we assumed that the field was only mildly
non-Gaussian and that we could take the unlensed temperature power
spectrum to calculate it. This is clearly not the case on small
scales. As we have shown in previous sections, on small scales the
fluctuations become very non-Gaussian as most of the power is
generated by lensing. We only consider modes with $l<3000$ for the
calculation of $C^{eff}_l$ to partially take into account this effect.

\subsection{The four point function}

In the first part of this  
paper we have studied one particular physical limit when we
are trying to recover information on the fluctuations of the mass
distribution on scales much smaller than the coherence length of
the CMB. To recover this limit we have to consider a quadrilateral
in which two sides are much smaller that the other two (figure
\ref{figquad}a). The two small sides correspond to the low pass
filtered derivatives while the large $\bi l$ correspond to the
high passed filtered one. We consider the case where $l_1,l_2 \ll
l_3\sim l_4$. As we noted before the power spectrum of the primary
anisotropies decreases exponentially while that of the deflection
angle is only as a power law. We conclude that of all the terms in
equation (\ref{fullthreefour}) only those explicitly written dominate,
\begin{eqnarray}
T_4(\bi l_1,\bi l_2,\bi l_3,\bi l_4)\approx 2 {\bi l_3 \cdot \bi
l_1\ \bi l_3 \cdot \bi l_2 \over l_3^2} C^{\delta\delta}_{l_3}
C^{\tl T \tl T}_{l_1} C^{\tl T \tl T}_{l_2},\ \ l_1,l_2 \ll
l_3\sim l_4,
\end{eqnarray}
where we approximated $C^{\delta\delta}_{l_3}\approx
C^{\delta\delta}_{l_4}$.

\begin{figure*}
\begin{center}
\leavevmode \epsfxsize=4.0in \epsfbox{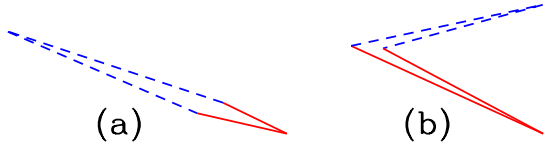}
\end{center}
\caption{Quadrilaterals corresponding to the two limits discussed
in the text. Panel a corresponds to configurations relevant for the
cross correlation between the large scale gradient and the small scale
power. Panel b shows the configurations that enter in the calculation
of $\cal S$ and $\cal E$.} \label{figquad}
\end{figure*}

A different set of quadrilaterals dominate in the calculation of $\cal
E$ and $\cal S$. Those variables extract information about the large
scale $\kappa$ fluctuations from small angular scale fluctuations in
the CMB. If we focus on modes of the temperature on scales larger that
the damping tail $l^2 C_l^{\tl T\tl T}$ remains approximately constant
while the power spectra of the deflection angle falls. Moreover 
$\cal S$ and $\cal E$ are combination of derivatives of the CMB and
the extra $l$s weigh the contribution to smaller scales. The power
spectra of $\cal S$ and $\cal E$ are dominated by the type of
quadrilaterals shown in figure \ref{figquad}b. All the $l$s are large
but the quadrilaterals are thin. The thin diagonal corresponds to the
$\bi l$ of the $\kappa$ mode being recovered. The terms in equation
(\ref{fullthreefour}) proportional to $C^{\tl
T \tl T}_{l_1} C^{\tl T \tl T}_{l_2} C^{\delta \delta}_{l_{12}}$
(where $l_1$ and $l_2$ represent the length of the sides and $l_{12}$
is the small diagonal) dominate. 

To extract all the information in the four point function 
we can add all the quadrilaterals with an appropriate weight,
\begin{eqnarray}
\hat Z&=&{A_f\over (2\pi)^2}
\int {d^2\bi l_1\over A_l}{d^2\bi l_2\over A_l}{d^2\bi l_3\over A_l}
\beta({\cal S})\ T(\bi l_1) T(\bi l_2)T(\bi l_3)T(\bi l_4), \label{defy}
\end{eqnarray}
where $\bi l_4=-(\bi l_1+\bi l_2+\bi l_3)$ and $A_l$ is the area in
$l$ space we are using. 
For the optimal filter that minimizes S/N one gets
$\beta \propto T_4/C^{\tl T \tl T}_{l1}C^{\tl T \tl T}_{l2}C^{\tl T
\tl T}_{l3}C^{\tl T \tl T}_{l4}$, where we have assumed  
Gaussianity to compute the variance.  
For the $S/N$ we get,
\begin{eqnarray}
({S\over N})^2&=&{A_f^{-1}\over 24(2\pi)^4}\int {d^2\bi l_1}
{d^2\bi l_2}{d^2\bi l_3}{T_4^2(\bi l_1,\bi l_2,\bi l_3,\bi l_4)
\over 
C_{l1}^{\tl T \tl T} C_{l2}^{\tl T \tl T} C_{l3}^{\tl T \tl T} 
C_{l4}^{\tl T \tl T}},\label{s2n1}
\end{eqnarray}
where the power spectra in the denominator should include the
contribution form detector noise. 

We change
integration variables in equation (\ref{s2n1}) and write,  
\begin{eqnarray}
({S\over N})^2&=&A_f^{-1}\int {d^2\bi l_1} ({4 C_{l1}^{\kappa \kappa}
\over \tl C^{eff}_l })^2 \nonumber \\
({1 \over \tl C^{eff}_l})^2 &=& {1\over 24}\int 
{d^2\bi l_2 \over (2\pi)^2 }{d^2\bi l_3 \over (2\pi)^2} {T_4^2(\bi l_1-\bi
l_2,\bi l_2,\bi l_3,-\bi l_1 - \bi l_3)
\over 
C_{\bi l1 - \bi l2}^{\tl T \tl T} C_{l2}^{\tl T \tl T} C_{l3}^{\tl T \tl T} 
C_{\bi l1 + \bi l3}^{\tl T \tl T} (4 C_{l1}^{\kappa \kappa})^2}.\label{s2n2}
\label{s2nnew4pt}
\end{eqnarray}
This is a useful change of variables because it makes the integral
resemble what we had in our old method. In this way 
the limit $l_1\rightarrow 0$ corresponds to quadrilaterals
that have four large sides but a small diagonal ($\bi l_1$). 
Equation (6) of \cite{psdmletter} reads,
\begin{equation}
({S\over N})^2=A_f^{-1}\int d^2 \bi l {(4 C_l^{\kappa \kappa})^2\over
\sigma_{C_l}^2},  
\label{s2nold4pt}
\end{equation}
with $\sigma_{C_l}^{-2}=\sigma^{-2}_{C^{\cal S \cal S}_l}+
\sigma^{-2}_{C^{\cal E \cal E}_l}+ \sigma^{-2}_{C^{\cal S \cal E}_l}$.

\begin{figure*}
\begin{center}
\leavevmode \epsfxsize=6.0in \epsfbox{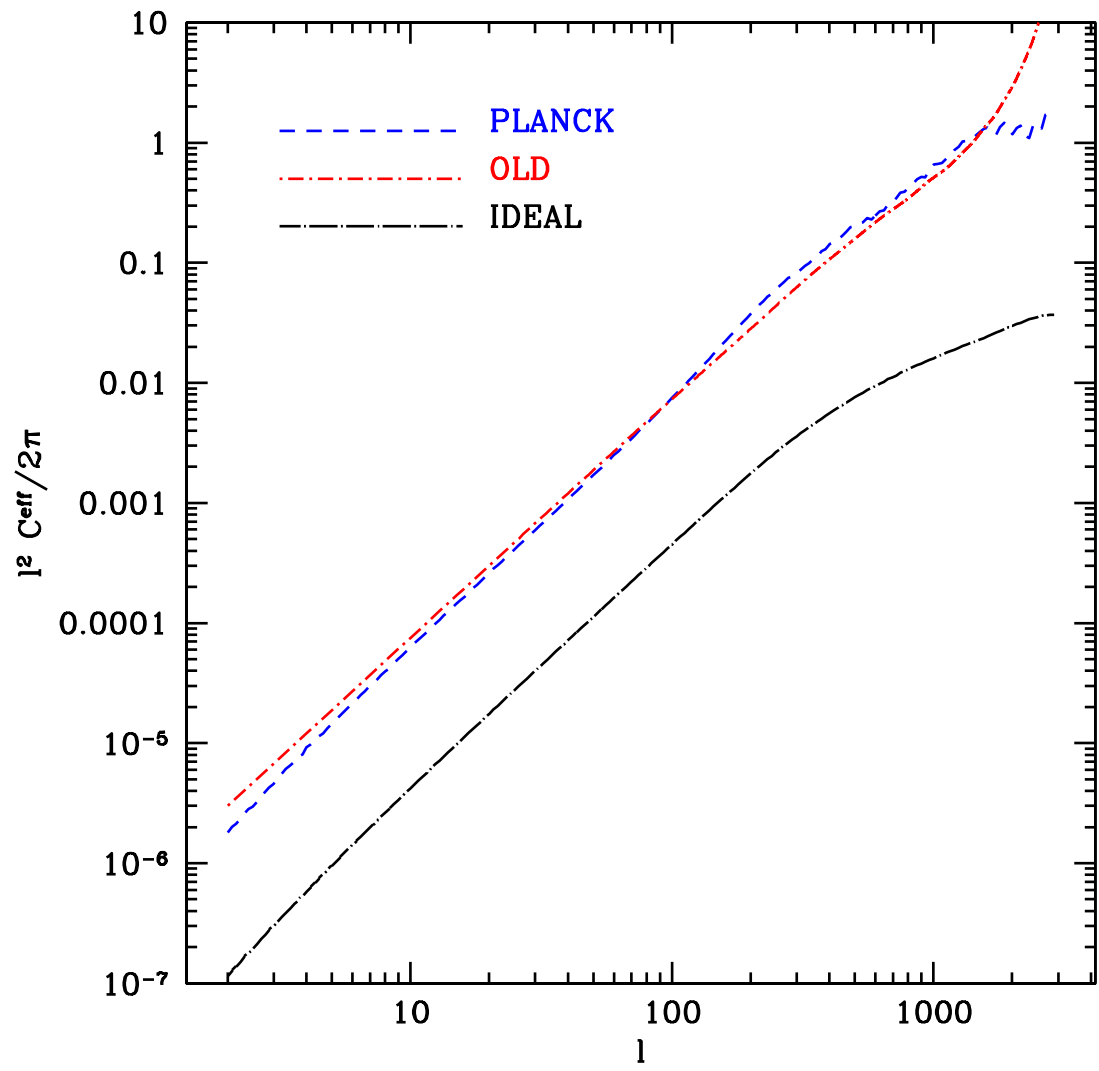}
\end{center}
\caption{Result of Monte Carlo for 4pt function together with result
from previous technique.}
\label{monte4pt}
\end{figure*}

In figure \ref{monte4pt} we compare the results  $C_l^{eff}$ in
(\ref{s2nnew4pt}) with $\sigma_{Cl}$ (\ref{s2nold4pt}). The plot is
qualitatively similar to figure \ref{comp3pt} for the three point function.  
While there are hardly any improvements in our previous method when we
go from Planck to an ideal experiment, there are significant
differences for the optimal filter which continues to
gather information from the damping tail. For Planck the situation is
different, both the optimal method and our old method obtain a
similar amount of information from the data. 
Even though the optimal method is able
to get information from the damping tail in the ideal case, 
this is unimportant
for Planck because the finite size of the beam makes it impossible. 
The fact that our previous method seems to have slightly less noise than the
optimal method when $l$ is a few hundred is most probably an
artifact. The noise in our old method was calculated using a Gaussian
approximation which se had seen braking down slightly in our
simulations \cite{longlens}.
As for the three point function we only included modes 
of the temperature with $l<3000$
as the power generated by lensing is non-Gaussian so our estimate of
the variance of $\hat Z$ is not valid on smaller scales. 

In the limit $l_1 \rightarrow 0$ the four point function becomes
approximately,
\begin{equation}
T_4(\bi l_1-\bi
l_2,\bi l_2,\bi l_3,-\bi l_1 - \bi l_3)\approx - C_{l2}^{\tl T \tl T}
C_{l3}^{\tl T \tl T} l^2_1 C_{l1}^{\delta \delta}.
\label{limitl120}
\end{equation}
To obtain equation (\ref{limitl120}) 
we had to assume that $C_l^{\kappa\kappa}$ is a
decreasing function of $l$ and that $C_{\bi l1 - \bi l2}^{\tl T \tl T}
\approx C_{l2}^{\tl T \tl T}$ and the equivalent formula for
$l_3$. Both of these assumptions break down in some range of
$l$s. For example, $C_l^{\kappa\kappa}$ has a peak at $l\sim 100$ 
and when $l_2$ is in the damping tale range, for finite $l_1$ there
might be corrections due to the difference between $C_{\bi l1 - \bi
l2}^{\tl T \tl T}$ and $C_{l2}^{\tl T \tl T}$.  
We can still use this expression as a rough estimate to try to
compare how the optimal $\beta$ compares with the weight used by our
previous method. In
this limit and for the quadrilaterals relevant for this statistic, the
optimal $\beta$ is equivalent to multiplying each of the temperatures
by $(C_l^{\tl T \tl T})^{1/2}/ 
(C_l^{\tl T \tl T} + B^2_l N_l^{\tl T \tl T})$. Thus for a temperature
power spectra that goes as $C_l^{\tl T \tl T}\propto l^{-2}$ and when
detector noise is irrelevant, the optimal filter amounts to
multiplying the temperatures by $l$, equivalent to taking
derivatives. This is the reason our previous method is not far from optimal in
situations where we can neglect detector noise and we are not trying
to extract information form the damping tail of the CMB. 
As we mention
when we discussed the 3pt function, on small enough scales our
treatment of the noise breaks down because the power is dominated by
the power generated by lensing. The Gaussian approximation for the
noise will not be valid. 


\section{Conclusions}

We have studied the generation of power by gravitational lensing
on small angular scales. We have shown that the power spectra of
the anisotropies gives a measure of the spectrum of the deflection
angle. The generated power is correlated with the size of the
large scale gradient.

The generation of power can be understood by studying the lensing
of the primary anisotropies by a cluster of galaxies. On the
scales of a cluster the CMB can be  assumed to be a simple
gradient. Lensing generates a wiggle on top of the gradient that
can be tens of $\mu K$. This signal will be large enough to be
detected by a CMB experiment which targeted clusters with
sufficient angular resolution, $\sim 1\ {\rm arcmin}$.

The lensing effect produced by the large scale structure of the
universe can be separated from other secondary effects or from
intrinsic CMB anisotropies at the last scattering surface by
measuring the cross correlation between the map of the large scale
gradient and the map of the small scale power. We have shown that
this statistic has only a slightly smaller  signal to noise than
the measurement of the small scale power itself. The power
generated by lensing dominates over the intrinsic fluctuations for
$l\succsim 4000$.

We have calculated the three and four point function of the lensing
field in the small angle limit. 
The cross correlation between large and small scales as well as
the statistics introduced in \cite{longlens} are particular
combinations of the four point function of the temperature field.
We have calculated explicitly the dependence of the three and four point
functions on the CMB and deflection angle power spectra as well as
on the shape of the configuration. It is fair to say that both the
statistic introduced here and those used in  \cite{longlens} can
be viewed as particular ways of compressing the information in the
four point function that take into account the physical intuition
coming from our understanding of the lensing effect. The lensing
effect predicts a particular dependence of the four point function
on configuration and scale that can be used to separate it from
other non Gaussian signals.

\smallskip
We are grateful to Uros Seljak and Wayne Hu for very 
very useful discussions.
M.Z. is supported by NASA through Hubble Fellowship grant
HF-01116.01-98A from STScI,
operated by AURA, Inc. under NASA contract NAS5-26555.


\begin{thebibliography}{99}

\bibitem[*]{urosemail} Electronic address: uros@mpa-garching.mpg.de
\bibitem[\ddagger]{matiasemail} Electronic address: matiasz@ias.edu

\bibitem{parameters}Jungman G., Kamionkowski M., Kosowsky A.,
and Spergel D. N. Phys. Rev. Lett., {\bf 76}, 1007 (1996);
{\it ibid} Phys. Rev. D {\bf 54}, 1332 (1996);
Bond J. R., Efstathiou G. 
\& Tegmark M., \MNRAS, {\bf 291}, 33 (1997);  
Zaldarriaga M., Spergel D. N. and Seljak U. 
aAstrophys. J. {\bf 488}, 1 (1997);  
Tegmark M., Eisenstein D., Hu W. and A. de Olivera Costa,
ApJ. {\bf 530}, 133 (2000)
\bibitem{zalrei}M. Zaldarriaga Phys. Rev. D {\bf 55}, 1822 (1997) 
\bibitem{bernardeau}Bernardeu F., Astron. \& Astrophys. {\bf 432},
15 (1997); Bernardeu F., Astron. \& Astrophys. {\bf 338} 767 (1998)
\bibitem{goldsper}Goldberg, D. M. and Spergel, D.N., astro-ph/9811251
\bibitem{psdmletter} U. Seljak and M. Zaldarriaga,  Phys. Rev.
Lett. {\bf 82}, 2636 (1999)
\bibitem{longlens} M. Zaldarriaga and U. Seljak,  Phys. Rev.
D {\bf 59}, 123507 (1999) 
\bibitem{metcalf} R. B. Metcalf and 
J. Silk, J. ApJ Lett.  {\bf 492}, L1 (1998) 
\bibitem{cbi} Information on CBI can be found at 
http://astro.caltech.edu/~tjp/CBI/abstract.html
\bibitem{uroslens} U. Seljak, ApJ. {\bf 463}, 1 (1996)
\bibitem{iswletter}  U. Seljak and M. Zaldarriaga,
Phys. Rev. D {\bf 60}, 043504 (1999) 
\bibitem{wayne} Cooray, A. and Hu, W. astro-ph/9910397 (ApJ in press)
\bibitem{uroszalcluster} U. Seljak and M. Zaldarriaga,
astro-ph/9907254 
\bibitem{knoxhaimanzal} L. Knox, Z. Haiman and M. Zaldarriaga, in
preparation.
\bibitem{pressetal}{\it Numerical Recepies}, W. Press et al.,
Cambridge University Press

\end{thebibliography}
\end{document}